\documentclass[aps,pre,twocolumn,groupedaddress,10pt,showpacs,superscriptaddress]{revtex4}
\usepackage{amsmath,amssymb}
\usepackage{graphicx,color}

\newcommand{\dif}{\mathrm{d}}%
\newcommand{\DiracDelta}[1]{\operatorname{\delta}\left(#1\right)}%

\newcommand{\ii}{i}%
\newcommand{\Laplace}{\boldsymbol{\triangle}}%
\newcommand{\Nabla}{\vec{\nabla}}%
\newcommand{\R}{\mathbb{R}}%

\begin{document}
\title{Polar liquid crystals in two spatial dimensions: 
the bridge from microscopic to macroscopic modeling}

\author{Raphael Wittkowski}\author{Hartmut L{\"o}wen}
\affiliation{Institut f{\"u}r Theoretische Physik II, Weiche Materie,
Heinrich-Heine-Universit{\"a}t D{\"u}sseldorf, D-40225 D{\"u}sseldorf, Germany}

\author{Helmut R. Brand}
\affiliation{Theoretische Physik III, Universit{\"a}t Bayreuth, D-95540 Bayreuth, Germany}
\date{\today}

\begin{abstract}
Two-dimensional polar liquid crystals have been discovered recently in monolayers of anisotropic molecules.
Here, we provide a systematic theoretical description of liquid-crystalline phases for polar particles in two spatial dimensions.
Starting from microscopic density functional theory, we derive a phase-field-crystal expression for the free-energy density
which involves three local order-parameter fields, namely the translational density, the polarization, and the nematic order parameter.
Various coupling terms between the order-parameter fields are obtained which are in line with macroscopic considerations. 
Since the coupling constants are brought into connection with the molecular correlations, we establish a bridge from microscopic
to macroscopic modeling.
Our theory provides a starting point for further numerical calculations of the stability of polar liquid-crystalline phases and is also relevant for modeling of microswimmers which are intrinsically polar.
\end{abstract}


\pacs{64.70.mf, 64.70.M-, 61.30.Dk}
\maketitle


\section{\label{sec:introduction}Introduction}
The study of liquid-crystalline phases formed by banana-shaped 
molecules opens the door to generate polar directions in a partially or 
completely fluid system due to a collective alignment of the polar axis 
of the banana-shaped (or bent-core) molecules \cite{BrandCP1992}.
So far, most of the liquid-crystalline phases formed by banana-shaped 
molecules are smectic \cite{NioriSWFT1996,LinkNSMCKW1997,BrandCP1998,BrandCP2003,PelzlDW1999}, 
but there have also been a few reports of nematic phases in this area 
\cite{PelzlDW1999,ShenDPWT1999,WeissflogNDPDEK2001,NioriYY2004}.
In parallel, there has been a considerable amount of work in Watanabe's group 
to generate polar nematic and cholesteric phases in liquid-crystalline polymers 
\cite{ChoiKPTNW1998,WatanabeHTWM1998,YenTPTW2006,KoikeEtAl2007,TaguchiYKTW2009}.
Among the polar nematic phases, a nematic phase with a symmetry as low as 
$C_{1\mathrm{h}}$ (or $C_{\mathrm{s}}$) was found \cite{KoikeEtAl2007} confirming 
earlier predictions about polar nematic phases with low symmetry \cite{BrandCP2000}. 

About 25 years ago, there has already been an early effort to synthesize polar 
nematics in systems composed of fairly large plate-like molecules \cite{HsiungRSSCK1987} 
(to avoid the flipping and thus to generate a lack of $\hat{n}\to-\hat{n}$ symmetry, 
with $\hat{n}$ being the average preferred direction usually called the 
\emph{director} \cite{deGennesP1995}). 
About the same time, compounds composed of pyramidic molecules were synthesized
with the same goal \cite{ZimmermannPLB1985}, but clear-cut evidence for a polar nematic could 
not be provided in either case.
This early work, however, triggered early modeling in the framework 
of a Ginzburg-Landau description \cite{PleinerB1989} and it was pointed out that 
phases with defects, in particular with spontaneous splay, should play an 
important role in such systems. It was predicted that a phase with defects 
would occur first in the vicinity of the phase transition to the polar nematic
phase.

In 2003 the group of Y.\ Tabe \cite{TabeYNYY2003} found a two-dimensional polar nematic phase
in Langmuir monolayers using the measurements of ferroelectric response and optical 
investigations in a low molecular weight compound composed of rod-like molecules. 
Very recently, there were two additional reports on a ferroelectric response
of a nematic phase in three-dimensional samples in compounds composed of 
bent-core molecules \cite{FrancescangeliEtAl2009,FrancescangeliS2010}, but it is open
whether the ferroelectric response was due to a field-induced reorganization
of cybotactic clusters -- as suggested by the authors -- or due to a 
bulk polar nematic behavior of a phase containing defects of the type outlined above.

Triggered by the reports of nematic phases in banana-shaped molecules, 
a macroscopic description of polar nematic phases in three spatial dimensions was 
derived \cite{BrandPZ2006,BrandCP2009}. 
It turned out that the absence of parity symmetry leads in such a fluid system 
to a number of cross-coupling terms between the macroscopic polarization and the 
other hydrodynamic variables, both statically and in the dissipative
dynamic regime.
In addition, it was found, both for reversible as well as for irreversible
dynamics, that there are new cross-coupling terms not present in typical 
liquid-crystalline systems not breaking parity symmetry, such as, for 
example, reversible dynamic cross-coupling terms between flow and temperature or 
concentration gradients.

Therefore, it is of high interest to have a more microscopic description evaluating 
the new cross-coupling terms quantitatively in order to aid synthesis of new materials 
for which corresponding effects can be substantial.
In this paper we start such a program using a phase-field-crystal (PFC) model 
\cite{ElderPBSG2007,Loewen2010,WittkowskiLB2010} to analyze the static behavior of 
polar phases in two spatial dimensions. This approach can be used as a bridge 
from microscopic to macroscopic modeling. 
We will systematically compare the results obtained from the PFC model to those obtained 
using symmetry based approaches such as the Ginzburg-Landau approach, a mean-field description 
of phase transitions neglecting fluctuations, and the approach of generalized hydrodynamics or 
macroscopic dynamics \cite{PleinerB1996}.

While in the former only variables are taken into account that lead to an infinite 
lifetime for excitations in the long wavelength limit, the approach of macroscopic dynamics 
also incorporates variables, which relax on a sufficiently long, but finite time scale 
in the limit of vanishing wave number. On realizing our program we strongly build on the
foundations given for the static PFC model for nematics and other phases with
orientational order in two \cite{Loewen2010} and three \cite{WittkowskiLB2010}
spatial dimensions.
In carrying out this program it turns out that it is of crucial importance 
for polar orientational order to go beyond the Ramakrishnan-Yussouff approximation 
\cite{RamakrishnanY1979}, which is usually used in the area of PFC models.
As a matter of fact many of the cross-coupling terms would not be obtained if
the Ramakrishnan-Yussouff approximation were implemented. 
The proposed model can be used as a starting point to explore phase transitions 
and interfaces for various polar liquid-crystalline sheets, in particular including 
plastic and full crystalline phases where the translational density shows a strong ordering.

The paper is organized as follows: in Sec.\ \ref{sec:derivation}, 
we derive a PFC model for polar liquid crystals.
Then, in Sec.\ \ref{sec:HB}, we discuss the relation of the two symmetry-based approaches 
with the PFC model studied in Sec.\ \ref{sec:derivation} and we show that many of the 
coefficients arising in the symmetry-based approaches can be linked to microscopic expressions 
via the PFC model.
We finally discuss possible extensions of the model to more complicated situations and give 
final conclusions in Sec.\ \ref{sec:conclusions}.

\section{\label{sec:derivation}Phase-field-crystal model for polar liquid crystals}
In general, a theory for polar liquid-crystalline phases can be constructed on three different levels.
First of all, a full \emph{microscopic} theory where the particle interactions and the thermodynamic 
conditions are the only input is provided by classical density functional theory (DFT)
\cite{Evans1979,Singh1991,Loewen1994a,Roth2010}.
DFT is typically used for isotropic particles 
\cite{RamakrishnanY1979,RosenfeldSLT1997,RothELK2002,HansenGoosM2009} but analogously holds for anisotropic particle interactions 
\cite{PoniewierskiH1988,GrafL1999,GrafL1998,HansenGoosM2009,FussnoteLoewen}.
The second level which may be called \emph{mesoscopic} is the phase-field approach 
where lowest-order gradients of an order-parameter field are considered \cite{Emmerich2008}.
This can be performed up to fourth-order gradients in order to describe a stable crystalline state with
order-parameter oscillations leading to the seminal PFC model of Elder and co-workers \cite{ElderKHG2002,ElderG2004,Emmerich2009}. 
The prefactors can be brought into relation with the microscopic DFT approach \cite{ElderPBSG2007,vanTeeffelenBVL2009}.
The PFC model has been extensively used to study numerically freezing and melting phenomena 
on microscopic length but much larger (diffusive) time scales 
\cite{TegzeGTPJANP2009,MellenthinKP2008,McKennaGV2009,HuangE2008,WuV2009,YuLV2009,JaatinenAEAN2009,JaatinenAN2010}. 
Finally, the third level are \emph{continuum} approaches \cite{deGennes1971,deGennes1973,deGennesP1995,ChaikinL1995,PleinerB1996} 
which respect the basic symmetries. Here, the prefactors are phenomenological elastic constants.
PFC modeling can be used to assign a microscopic meaning to the prefactors thus linking the 
microscopic DFT approach to the symmetry-based approach.

In this section, a PFC model for polar liquid crystals in two spatial dimensions is derived from DFT by a systematic gradient expansion
of various coarse-grained order-parameter fields. As a result, we get a free-energy functional which involves 
the order-parameter fields and their spatial derivatives.
The prefactors of various contributions are expressed as generalized moments of direct correlation functions 
in the isotropic state which provides a bridge between microscopic density functional theory and macroscopic approaches.

\subsection{Static free-energy functional}
We consider a two-dimensional system of $N$ anisotropic particles with the center-of-mass 
positions $\vec{r}_{i}$ and orientations that are characterized by the unit vectors $\hat{u}_{i}$ with $i\in\{1,\dotsc,N\}$.
To provide uniaxiality, we assume the existence of a symmetry axis for the anisotropic particles. 
Furthermore, we assume a broken head-tail-symmetry, i.\ e., we assume polar particles. This polar system is restricted 
to the domain $\mathcal{A}\subseteq\R^{2}$ with the total area 
\begin{equation}
A=\int_{\mathcal{A}}\!\!\!\dif\vec{r}
\end{equation}
and kept at a finite temperature $T$.
The polar particles are supposed to interact in accordance with a prescribed pair-interaction potential 
$\operatorname{V}(\vec{r}_{1}-\vec{r}_{2},\hat{u}_{1},\hat{u}_{2})$. Typical examples include particles with an embedded
dipole moment \cite{LombaLW2000,FroltsovBLL2003,AlvarezMW2008} modeled by a dipolar hard disk potential, colloidal
pear-like particles \cite{KegelBEP2006,HoseinJLEL2009} with corresponding excluded volume interactions, 
Janus particles \cite{HongCLG2006,HoCSLK2008} which possess two different sides, and asymmetric brush polymers 
modeled by Gaussian segment potentials \cite{RexWL2007}.

We define the one-particle density field as
\begin{equation}
\rho(\vec{r},\hat{u}) = \bigg\langle\sum^{N}_{i=1} 
\DiracDelta{\vec{r}-\vec{r}_{i}} \DiracDelta{\hat{u}-\hat{u}_{i}}\bigg\rangle 
\end{equation}
with the mean particle number density 
\begin{equation}
\bar{\rho}=\frac{N}{A}\;,
\end{equation}
where
\begin{equation}
\langle\mathfrak{O}\rangle = \frac{1}{\mathcal{Z}} \int_{\mathcal{A}}\!\!\!\dif^{N}\vec{r}
\int_{\mathcal{S}_{1}}\!\!\!\!\dif^{N}\hat{u} \,\, \mathfrak{O} \,
e^{-\beta\!\sum^{N}_{i<j=1} \operatorname{V}(\vec{r}_{i}-\vec{r}_{j},\hat{u}_{i},\hat{u}_{j})}
\end{equation}
is the classical canonical average of the observable $\mathfrak{O}$. Here, 
we introduced the notation $\dif^{n}\vec{x}=\dif\vec{x}_{1}\dotsb\dif\vec{x}_{n}$ for an arbitrary vector $\vec{x}$ and $n\in\mathbb{N}$. 
$\mathcal{Z}$ denotes the classical canonical partition function and guarantees correct normalization such that $\langle1\rangle=1$. 
Furthermore, $\beta=1/(k_{\mathrm{B}}T)$ is the inverse temperature with the Boltzmann constant $k_{\mathrm{B}}$  
and $\mathcal{S}_{1}$ is the unit circle. 
The one-particle density $\rho(\vec{r},\hat{u})$ describes the probability density $\rho(\vec{r},\hat{u})/\bar{\rho}$ to find a particle 
with orientation $\hat{u}$ at position $\vec{r}$. Due to the restriction on two spatial dimensions, 
the orientation $\hat{u}(\varphi)=(\cos(\varphi),\sin(\varphi))$ is entirely defined by the polar angle $\varphi$. 
A collective ordering of a set of particles may lead to a macroscopic polarization whose local direction can be expressed by the 
space-dependent dimensionless unit vector $\hat{p}(\vec{r})=\hat{u}(\varphi_{0}(\vec{r}))$, that is parametrized by a scalar 
order-parameter field $\varphi_{0}(\vec{r})$.

Under the assumption of small anisotropies in the orientation, it is now possible to expand the one-particle density $\rho(\vec{r},\hat{u})$ 
with respect to the angle $\varphi-\varphi_{0}(\vec{r})$ between the particular orientation $\hat{u}$ and the macroscopic 
polarization $\hat{p}(\vec{r})$ into a Fourier series. 
Throughout this paper we will assume explicitly that the preferred direction associated with dipolar order, $\hat{p}$,   
and the direction associated with quadrupolar order, $\hat{n}$, are parallel.
We will therefore use $\hat{p}$ in the following.
In general, these two types of order can be associated with two different preferred directions (compare, e.\ g., reference \cite{BrandCP2000}).
The expansion with respect to orientation results in the approximation 
\begin{equation}
\begin{split}
\rho(\vec{r},\hat{u}) \approx 
\bar{\rho}\,\Big(1 &+ \psi_{1}(\vec{r}) 
+ P(\vec{r})\big(\hat{p}(\vec{r})\cdot\hat{u}\big) \\
&+ S(\vec{r})\Big(\big(\hat{p}(\vec{r})\cdot\hat{u}\big)^{2}-\frac{1}{2}\Big)\Big) \;,
\end{split}
\label{eq:rho}
\end{equation}
where the Fourier series is truncated at second order.
Here, we introduced three additional dimensionless order-parameter fields $\psi_{1}(\vec{r})$, $P(\vec{r})$, and $S(\vec{r})$.
These order-parameter fields are the reduced orientationally averaged translational density 
\begin{equation}
\psi_{1}(\vec{r}) = \frac{1}{2\pi\bar{\rho}}\int_{\mathcal{S}_{1}}\!\!\!\!\dif\hat{u}
\,\big(\rho(\vec{r},\hat{u})-\bar{\rho}\big) \;, 
\label{eq:psi1p}
\end{equation}
the strength of the polarization 
\begin{equation}
P(\vec{r}) = \frac{1}{\pi\bar{\rho}}\int_{\mathcal{S}_{1}} \!\!\!\!\dif\hat{u} 
\,\rho(\vec{r},\hat{u}) \big(\hat{p}(\vec{r})\cdot\hat{u}\big)
\label{eq:Pp}
\end{equation}
and the nematic order parameter 
\begin{equation}
S(\vec{r}) = \frac{4}{\pi\bar{\rho}}\int_{\mathcal{S}_{1}}\!\!\!\!\dif\hat{u} 
\,\rho(\vec{r},\hat{u})\Big(\big(\hat{p}(\vec{r})\cdot\hat{u}\big)^{2}-\frac{1}{2}\Big) 
\label{eq:psi2p}
\end{equation}
that measures the local degree of orientational order. 
The strength $P(\vec{r})$ of the polarization and the director $\hat{p}(\vec{r})$ are modulus and orientation of the polarization 
$\vec{P}(\vec{r})=P(\vec{r})\hat{p}(\vec{r})$.
Note that for apolar particles \cite{Loewen2010} $P(\vec{r})=0$ such that apolar particles result as a special limit from the present theory. 

Now we refer to microscopic density functional theory which is typically formulated 
for spherical systems \cite{Evans1979,Singh1991,Loewen1994a} but can also be constructed 
for anisotropic particle interactions (which dates back to Onsager) 
\cite{PoniewierskiH1988,GrafL1999,GrafL1998,HansenGoosM2009,FussnoteLoewen}.
Density functional theory establishes the existence of a free-energy functional
$\mathcal{F}[\rho(\vec{r},\hat{u})]$ of the one-particle density $\rho(\vec{r},\hat{u})$ which becomes minimal
for the equilibrium density. The total functional can be split into an
ideal rotator gas functional and an excess functional:
\begin{equation}
\mathcal{F}[\rho(\vec{r},\hat{u})] = \mathcal{F}_{\textrm{id}}[\rho(\vec{r},\hat{u})] 
+ \mathcal{F}_{\textrm{exc}}[\rho(\vec{r},\hat{u})] \;. 
\end{equation}
The ideal gas functional is local and nonlinear, it is exactly given by 
\begin{equation}
\beta\mathcal{F}_{\textrm{id}}[\rho(\vec{r},\hat{u})] =\! 
\int_{\mathcal{A}}\!\!\!\dif\vec{r}\int_{\mathcal{S}_{1}} 
\!\!\!\!\dif\hat{u}\, \rho(\vec{r},\hat{u}) \big(\ln(\Lambda^{2}\rho(\vec{r},\hat{u}))-1\big) 
\label{eq:FidGp}
\end{equation}
where $\Lambda$ denotes the thermal de-Broglie-wavelength.
The excess functional $\mathcal{F}_{\textrm{exc}}[\rho(\vec{r},\hat{u})]$, on the other hand, is in general
(i.\ e., for a non-vanishing $\operatorname{V}(\vec{r}_{1}-\vec{r}_{2},\hat{u}_{1},\hat{u}_{2})$) unknown
and approximations are needed. However, there is a formally exact expression gained from a functional Taylor
expansion in the density variations $\Delta\rho(\vec{r},\hat{u})=\rho(\vec{r},\hat{u})-\bar{\rho}$ 
around a homogeneous reference density $\bar{\rho}$ \cite{Evans1979}:
\begin{equation}
\beta\mathcal{F}_{\mathrm{exc}}[\rho(\vec{r},\hat{u})] = \beta\mathcal{F}^{(0)}_{\mathrm{exc}}(\bar{\rho})
-\sum^{\infty}_{n=2}\frac{1}{n!}\mathcal{F}^{(n)}_{\mathrm{exc}}[\rho(\vec{r},\hat{u})] 
\label{eq:FTE}
\end{equation}
with the $n$-th order contributions 
\begin{equation}
\mathcal{F}^{(n)}_{\mathrm{exc}}[\rho(\vec{r},\hat{u})] = \!
\int_{\mathcal{A}} \!\!\!\dif\underline{\vec{r}} \int_{\mathcal{S}_{1}} \!\!\!\!\dif\underline{\hat{u}} \,
c^{(n)}(\underline{\vec{r}},\underline{\hat{u}}) \prod^{n}_{i=1}\Delta\rho(\vec{r}_{i},\hat{u}_{i}) \;.
\label{eq:TET}
\end{equation}
Here, $c^{(n)}(\underline{\vec{r}},\underline{\hat{u}})$ denotes the $n$-particle direct correlation function, 
and the notation $\underline{\vec{x}}=(\vec{x}_{1},\dotsc,\vec{x}_{n})$ for an arbitrary vector $\vec{x}$ is used.
The first term on the right-hand side of Eq.\ \eqref{eq:FTE} corresponds to $n=0$ and is an irrelevant 
constant that can be neglected. We remark that also the first-order term ($n=1$ in Eq.\ \eqref{eq:TET}) vanishes
since in a homogeneous reference state $c^{(1)}(\vec{r}_{1},\hat{u}_{1})$ must be constant due to
translational and orientational symmetry.

For isotropic particles, various approximations based on expression \eqref{eq:FTE} have been proposed.
The theory of Ra\-ma\-krish\-nan and Yus\-souff  \cite{RamakrishnanY1979} keeps only second-order 
terms in the expansion. This provides a microscopic theory for freezing both in three
\cite{RamakrishnanY1979} and two spatial dimensions \cite{vanTeeffelenLHL2006}. More refined
approaches include the third-order term \cite{Barrat1987} with an
approximate triplet direct correlation function \cite{BarratHP1987,BarratHP1988}, but a 
perturbative fourth-order theory has never been considered. Complementary, non-perturbative approaches
like the recently proposed fundamental-measure theory for arbitrarily shaped hard particles 
\cite{HansenGoosM2009} include direct correlation functions of arbitrary order.

We now insert the parametrization \eqref{eq:rho} of the one-particle density into Eqs.\ \eqref{eq:FidGp} and \eqref{eq:FTE} 
in order to obtain a free-energy functional of the order-parameter fields $\psi_{1}(\vec{r})$, $P(\vec{r})$, $S(\vec{r})$, 
and $\hat{p}(\vec{r})$.
First, after inserting the density parameterization \eqref{eq:rho} into the ideal gas functional \eqref{eq:FidGp},
we expand the logarithm and truncate the expansion of the integrand at fourth order.
This order guarantees stabilization of the solutions (similar to the traditional Ginzburg-Landau theory of phase transitions).
Performing the angular integration results in the approximation 
\begin{equation}
\beta\mathcal{F}_{\textrm{id}}[\rho(\vec{r},\hat{u})] \approx F_{\textrm{id}} 
+ \pi\bar{\rho}\int_{\mathcal{A}} \!\!\!\dif\vec{r} \,f_{\mathrm{id}}
\label{eq:Fidpa}
\end{equation}
with the local ideal rotator gas free-energy density 
\begin{equation}
\begin{aligned}
&f_{\mathrm{id}} = 
2\:\!\psi_{1} + \psi^{2}_{1} - \frac{\psi^{3}_{1}}{3} + \frac{\psi^{4}_{1}}{6} + \frac{P^{2}}{2} - \frac{\psi_{1}P^{2}}{2} \\ 
&\qquad\quad\;\; 
+ \frac{\psi^{2}_{1}P^{2}}{2} 
- \frac{P^{2}S}{8} + \frac{\psi_{1}P^{2}S}{4} 
+ \frac{P^{4}}{16} \\[2pt] 
&\qquad\quad\;\; 
+ \frac{S^{2}}{8} - \frac{\psi_{1}S^{2}}{8} 
+ \frac{\psi^{2}_{1}S^{2}}{8} 
+ \frac{P^{2}S^{2}}{16} 
+ \frac{S^{4}}{256}
\end{aligned}
\label{eq:Fidpb}
\end{equation}
and the abbreviation
\begin{equation}
F_{\textrm{id}} = 2\pi\bar{\rho}\,A\,(\ln(\Lambda^{2}\bar{\rho})-1)
\label{eq:FNp}
\end{equation}
for a constant and therefore irrelevant term.

Secondly, we insert the density parametrization \eqref{eq:rho} into Eq.\ \eqref{eq:FTE}.
We will truncate this expansion at fourth order.
Since the $n$-th order direct correlation function $c^{(n)}$ in Eq.\ \eqref{eq:FTE} is not known in general, 
we expand it into a Fourier series with respect to its orientational degrees of freedom.
By considering the translational and rotational invariance of the direct correlation function, we can use the parametrization 
$c^{(n+1)}(\underline{R},\underline{\phi_{\mathrm{R}}},\underline{\phi})$ with $\underline{R}=(R_{1},\dotsc,R_{n})$, 
$\underline{\phi_{\mathrm{R}}}=(\phi_{\mathrm{R}_{1}},\dotsc,\phi_{\mathrm{R}_{n}})$, and $\underline{\phi}=(\phi_{1},\dotsc,\phi_{n})$
for the direct correlation function $c^{(n+1)}$ to reduce its orientational degrees of freedoms from $2n+2$ to $2n$.  
Here, the new variables are related to the previous ones by 
$\vec{r}_{1}-\vec{r}_{i+1}=R_{i}\hat{u}(\varphi_{\mathrm{R}_{i}})$, 
$\hat{u}_{i}=\hat{u}(\varphi_{i})$, 
$\phi_{\mathrm{R}_{i}}=\varphi_{1}-\varphi_{\mathrm{R}_{i}}$, and
$\phi_{i}=\varphi_{1}-\varphi_{i+1}$.
With this parametrization, the Fourier expansion of the direct correlation function reads
\begin{equation}
c^{(n+1)}(\underline{R},\underline{\phi_{\mathrm{R}}},\underline{\phi}) = \!\!\!
\sum^{\infty}_{\begin{subarray}{c}l_{j},m_{j}=-\infty\\1\leqslant j\leqslant n\end{subarray}}\!\!\!\!\! 
\tilde{c}^{(n+1)}_{\underline{l},\underline{m}}
\!(\underline{R}) e^{\ii(\underline{l}\cdot\underline{\phi_{\mathrm{R}}} + \underline{m}\cdot\underline{\phi})} 
\label{eq:cexp}
\end{equation}
with the expansion coefficients 
\begin{equation}
\begin{split}
&\tilde{c}^{(n+1)}_{\underline{l},\underline{m}}
\!(\underline{R}) = \frac{1}{(2\pi)^{2n}}
\int^{2\pi}_{0}\!\!\!\!\!\!\dif\underline{\phi_{\mathrm{R}}}
\int^{2\pi}_{0}\!\!\!\!\!\!\dif\underline{\phi} \\
&\qquad \times c^{(n+1)}(\underline{R},\underline{\phi_{\mathrm{R}}},\underline{\phi})
e^{-\ii(\underline{l}\cdot\underline{\phi_{\mathrm{R}}} + \underline{m}\cdot\underline{\phi})} \;.
\end{split}
\label{eq:ec}
\end{equation}
Next, we set $\mathcal{A}=\R^{2}$ and perform a gradient expansion \cite{LoewenBW1989,LoewenBW1990,OhnesorgeLW1991,Lutsko2006,ElderPBSG2007} 
in the order-parameter fields.
For the term \eqref{eq:TET} corresponding to $n=2$, this gradient expansion is performed up to fourth order in $\psi^{2}_{1}(\vec{r})$ 
to allow stable crystalline phases and up to second order in all other order-parameter products, where we assume that the highest-order
gradient terms ensure stability.
However, for $n=3$ and $n=4$ we truncate the gradient expansion at first and zeroth order, respectively.
This results in the components 
\begin{equation}
\mathcal{F}^{(n)}_{\mathrm{exc}}[\psi_{1},P,S,\hat{p}] \approx\! 
\int_{\R^{2}} \:\!\!\!\!\!\!\dif\vec{r} \, f^{(n)}_{\mathrm{exc}}
\end{equation}
of the static excess free-energy functional.
In this equation, the excess free-energy densities $f^{(n)}_{\mathrm{exc}}(\vec{r})$ are local and given by
\begin{widetext}
\allowdisplaybreaks
\begin{align}
\begin{split}
&f^{(2)}_{\mathrm{exc}} = 
A_{1}\psi^{2}_{1} + A_{2}\big(\Nabla\psi_{1}\big)^{2} + A_{3}\big(\Laplace\psi_{1}\big)^{2} 
+B_{1}\psi_{1}\Nabla\cdot\big(\hat{p}P\big) + B_{2}S\Big(\hat{p}\cdot\Nabla P - P\big(\Nabla\cdot\hat{p}\big)\Big) \\ 
&\qquad\qquad+ B_{3} \Big( 
\Nabla\psi_{1}\cdot\Nabla S 
-2\big(\hat{p}\cdot\Nabla\psi_{1}\big) 
\big(\hat{p}\cdot\Nabla S\big) -2 S\Nabla\psi_{1}\cdot 
\big( \big(\hat{p}\cdot\Nabla\big)\hat{p} + 
\hat{p}\big(\Nabla\cdot\hat{p}\big) \big)\Big) \\ 
&\qquad\qquad+ P^{2} 
\Big( C_{1} - C_{2} 
\big(\hat{p}\cdot\Laplace\hat{p}\big) 
-C_{3} 
\big(\hat{p}\cdot\Nabla\big) \big(\Nabla\cdot\hat{p}\big) \Big) 
+ C_{2} \big(\Nabla P\big)^{2} 
+ C_{3}\big(\hat{p}\cdot\Nabla P\big)^{2} \\
&\qquad\qquad+ S^{2}\Big(D_{1} - 4D_{2}\big(\hat{p}\cdot\Laplace\hat{p}\big)\Big) +D_{2}\big(\Nabla S\big)^{2} \;,
\end{split}\label{eq:Fexc_GpEa}\\[2mm]
\begin{split}
&f^{(3)}_{\mathrm{exc}} =
E_{1}\psi^{3}_{1} + E_{2}\psi_{1}P^{2} + E_{3}\psi_{1}S^{2} + E_{4}S P^{2} 
+\big(F_{1}\psi_{1}+F_{2}S\big)P\big(\hat{p}\cdot\Nabla\psi_{1}\big) \\
&\qquad\qquad+ \big(2F_{3}\psi_{1}S+F_{4}P^{2}+F_{5}S^{2}\big)\big(\hat{p}\cdot\Nabla P\big)
+\big(F_{3}\psi_{1}+F_{6}S\big)P\big(\hat{p}\cdot\Nabla S\big) \;,
\end{split}\label{eq:Fexc_GpEb}\\[2mm]
\begin{split}
&f^{(4)}_{\mathrm{exc}} =
G_{1}\psi^{4}_{1} + G_{2}\psi^{2}_{1}P^{2} + G_{3}\psi^{2}_{1}S^{2} + G_{4}\psi_{1}P^{2}S
+ G_{5}P^{2}S^{2} + G_{6}P^{4} + G_{7}S^{4}
\end{split}\label{eq:Fexc_GpEc}
\end{align}%
\end{widetext}
with the coefficients 
\begin{equation}%
A_{1} = 8\,\mathrm{M}^{0}_{0}(1) \,,\quad A_{2} = -2\,\mathrm{M}^{0}_{0}(3) \,,\quad
A_{3} = \frac{1}{8}\,\mathrm{M}^{0}_{0}(5)
\label{eq:Ca}
\end{equation}%
in the gradient expansion in $\psi^{2}_{1}(\vec{r})$, that also appear -- in a different form -- in the 
traditional PFC model of Elder and co-workers \cite{ElderPBSG2007}. The coefficients
{\allowdisplaybreaks\begin{align}%
B_{1} &= 4\Big(\mathrm{M}^{0}_{1}(2)-\mathrm{M}^{1}_{-1}(2)\Big) \;, \\
B_{2} &= \mathrm{M}^{2}_{-1}(2)-\mathrm{M}^{1}_{1}(2) \;, \\
B_{3} &= \frac{1}{2}\Big(\mathrm{M}^{2}_{-2}(3)+\mathrm{M}^{0}_{2}(3)\Big)
\label{eq:Cb}
\end{align}}%
belong to the terms that contain gradients and the modulus of the polarization $P(\vec{r})$ in first order or that describe the coupling 
between gradients in the translational density $\psi_{1}(\vec{r})$ and gradients in the nematic order parameter $S(\vec{r})$, respectively.
The following three coefficients 
{\allowdisplaybreaks\begin{align}%
C_{1} &= 4\,\mathrm{M}^{1}_{0}(1) \;, \\
C_{2} &= \frac{1}{2}\,\mathrm{M}^{1}_{-2}(3) - \mathrm{M}^{1}_{0}(3) \;, \\
C_{3} &= -\mathrm{M}^{1}_{-2}(3)
\label{eq:Cc}
\end{align}}%
appear in the gradient expansion regarding $P^{2}(\vec{r})$ and    
\begin{equation}%
D_{1} = \mathrm{M}^{2}_{0}(1) \,,\qquad D_{2} = -\frac{1}{4}\,\mathrm{M}^{2}_{0}(3) 
\label{eq:Cd}
\end{equation}%
are the coefficients of the gradient expansion in $S^{2}(\vec{r})$. So far, all these coefficients can also be obtained
by using the second-order Ramakrishnan-Yussouff functional for the excess free energy. 
The remaining coefficients, however, result from higher-order contributions in our functional Taylor expansion. 
In third order, we find for the homogeneous terms the coefficients 
{\allowdisplaybreaks\begin{align}%
E_{1} &= 32\,\widehat{\mathrm{M}}^{00}_{00} \;, \\
E_{2} &= 16\Big(\widehat{\mathrm{M}}^{-11}_{00}+2\,\widehat{\mathrm{M}}^{01}_{00}\Big) \;, \\
E_{3} &= 4\Big(\widehat{\mathrm{M}}^{-22}_{00}+2\,\widehat{\mathrm{M}}^{02}_{00}\Big) \;, \\
E_{4} &= 4\Big(2\,\widehat{\mathrm{M}}^{-21}_{00}+\widehat{\mathrm{M}}^{11}_{00}\Big)
\label{eq:Ce}
\end{align}}%
and for the terms containing a gradient we find the coefficients
{\allowdisplaybreaks\begin{align}%
F_{1} &= -32\Big(\widetilde{\mathrm{M}}^{-10}_{01}-2\,\widetilde{\mathrm{M}}^{0-1}_{01}+\widetilde{\mathrm{M}}^{00}_{01}\Big) \;, \\
F_{2} &= -8\Big(\widetilde{\mathrm{M}}^{-20}_{01}+\widetilde{\mathrm{M}}^{-21}_{01}-2\,\widetilde{\mathrm{M}}^{0-2}_{01}
-2\,\widetilde{\mathrm{M}}^{1-2}_{01} \\
&\qquad\quad+\widetilde{\mathrm{M}}^{10}_{01}+\widetilde{\mathrm{M}}^{01}_{01}\Big) \;, \\
F_{3} &= -8\Big(\widetilde{\mathrm{M}}^{-21}_{01}-\widetilde{\mathrm{M}}^{0-2}_{01}-\widetilde{\mathrm{M}}^{1-2}_{01}
+\widetilde{\mathrm{M}}^{01}_{01}\Big) \;, \\
F_{4} &= 16\Big(\widetilde{\mathrm{M}}^{-1-1}_{01}-2\,\widetilde{\mathrm{M}}^{-11}_{01}+\widetilde{\mathrm{M}}^{1-1}_{01}\Big) \;, \\
F_{5} &= -4\Big(\widetilde{\mathrm{M}}^{-22}_{01}-\widetilde{\mathrm{M}}^{-1-2}_{01}+\widetilde{\mathrm{M}}^{-12}_{01}
-\widetilde{\mathrm{M}}^{2-2}_{01}\Big) \;, \\
F_{6} &= 2\Big(2\,\widetilde{\mathrm{M}}^{-2-1}_{01}-5\,\widetilde{\mathrm{M}}^{-22}_{01}-5\widetilde{\mathrm{M}}^{-12}_{01}
+3\,\widetilde{\mathrm{M}}^{-1-2}_{01} \\
&\qquad\;\!+3\,\widetilde{\mathrm{M}}^{2-2}_{01}+2\,\widetilde{\mathrm{M}}^{2-1}_{01}\Big) \;.
\label{eq:Cf}
\end{align}}%
In fourth order, we only kept homogeneous terms. The corresponding coefficients are
{\allowdisplaybreaks\begin{align}%
G_{1} &= 128\,\widehat{\mathrm{M}}^{000}_{000} \;, \\
G_{2} &= 192\Big(\widehat{\mathrm{M}}^{-101}_{000}+\widehat{\mathrm{M}}^{001}_{000}\Big) \;, \\
G_{3} &= 48\Big(\widehat{\mathrm{M}}^{-202}_{000}+\widehat{\mathrm{M}}^{002}_{000}\Big) \;, \\
G_{4} &= 48\Big(2\,\widehat{\mathrm{M}}^{-201}_{000}+\widehat{\mathrm{M}}^{-211}_{000}+\widehat{\mathrm{M}}^{011}_{000}\Big) \;, \\
G_{5} &= 24\Big(\widehat{\mathrm{M}}^{-212}_{000}+\widehat{\mathrm{M}}^{-112}_{000}\Big) \;, \\
G_{6} &= 48\,\widehat{\mathrm{M}}^{-111}_{000} \;, \\
G_{7} &= 3\,\widehat{\mathrm{M}}^{-222}_{000} \;.
\label{eq:Cg}
\end{align}}%
All the coefficients from above are linear combinations of moments of the direct correlation functions. 
These moments are defined through  
\begin{equation}
\mathrm{M}^{\underline{m}}_{\underline{l}}(\underline{\alpha}) = 
\pi^{2n+1}\bar{\rho}^{n+1} \Bigg(\prod^{n}_{i=1} \int^{\infty}_{0}\!\!\!\!\!\!\dif R_{i}R^{\alpha_{i}}_{i}\Bigg) 
\tilde{c}^{(n+1)}_{\underline{l},\underline{m}}(\underline{R}) \;.
\label{eq:M}
\end{equation}
To shorten the notation, we introduced the abbreviations 
$\widehat{\mathrm{M}}^{\underline{m}}_{\underline{l}}=\mathrm{M}^{\underline{m}}_{\underline{l}}(\underline{1})$ and
$\widetilde{\mathrm{M}}^{m_{1}m_{2}}_{l_{1}l_{2}}=\mathrm{M}^{m_{1}m_{2}}_{l_{1}l_{2}}(1,2)$ and used some symmetry considerations
that are outlined in appendix \ref{Anhang:SC}.
The moments over expansion coefficients of the direct correlation functions depend on the particular thermodynamic conditions 
and therefore on the parameters $\bar{\rho}$ and $T$. 

For stability reasons, we assume that the coefficients of the highest-order terms in the gradients and order-parameter fields 
are positive in the full free-energy functional. If this appears not to be the case for a certain system, 
it is necessary to take into account further terms of the respective order-parameter field up to the first stabilizing order. 
  
Eqs.\ \eqref{eq:Fexc_GpEa}-\eqref{eq:Fexc_GpEc} constitute the main result of the paper: it is a systematic gradient expansion of 
order-parameter fields in the free-energy functional. The prefactors are moments of various 
direct correlation functions and therefore provide the link towards microscopic correlations. This is similar in 
spirit to PFC models \cite{ElderPBSG2007,ElderG2004,JaatinenAN2010,TegzeGTPJANP2009,Emmerich2009,vanTeeffelenBVL2009} 
for spherical particles.

\subsection{Special cases of the phase-field-crystal model}
We now discuss special cases of our model. First of all, Eqs.\ \eqref{eq:Fexc_GpEa}-\eqref{eq:Fexc_GpEc} are an extension of the excess free-energy density for apolar particles, that was recently proposed in reference \cite{Loewen2010}. 
This extension comprises a possible polarization of liquid-crystalline particles as well as an enlarged functional Taylor expansion that is beyond the scope of the second-order (Ramakrishnan-Yussouff) approximation. 
Because of that, our free-energy functional contains a few simpler models as special cases and is therefore the main result of this paper. 
Two special models that are known from literature and can be obtained from our model by setting some of the order-parameter fields to zero are the traditional PFC model of Elder and co-workers \cite{ElderPBSG2007} for isotropic particles without orientational degrees of freedom and the PFC model of L{\"o}wen \cite{Loewen2010} for apolar anisotropic liquid crystals in two spatial dimensions.
In comparison with our free-energy functional, the two mentioned models base on the Ramakrishnan-Yussouff approximation. 
The traditional PFC model has only one order-parameter field. This is the translational density which corresponds to $\psi_{1}(\vec{r})$ in our model. 
If we set all order-parameter fields that are related to orientational degrees of freedom in our PFC model to zero, i.\ e., 
$P(\vec{r})=0$ and $S(\vec{r})=0$, and neglect all higher-order contributions for $n\geqslant 2$ in the functional Taylor expansion \eqref{eq:FTE}, 
then we obtain the traditional PFC model of Elder and co-workers.
The PFC model of L{\"o}wen considers anisotropic particles with one orientational degree of freedom but no polarization. 
Therefore, this PFC model results from our model for a vanishing polarization $P(\vec{r})=0$.
Also here, we have to neglect all contributions \eqref{eq:TET} for $n\geqslant 2$.

\section{\label{sec:HB}Macroscopic approaches}
In this section, we investigate the bridge between the PFC model presented in detail in the last section for 
polar liquid crystals in two spatial dimensions and the symmetry-based macroscopic approaches:
\emph{Ginzburg-Landau} and \emph{generalized continuum} description. 
In addition, we can also compare these results obtained for polar liquid crystals in two spatial dimensions with 
those obtained previously for non-polar liquid crystals in two \cite{Loewen2010} as well as in three 
\cite{WittkowskiLB2010} spatial dimensions.

The general PFC results of this paper have been summarized in Eqs.\ \eqref{eq:Fexc_GpEa}-\eqref{eq:Fexc_GpEc}.
We first analyze the terms given in Eq.\ \eqref{eq:Fexc_GpEa}, which are quadratic in the variables and their 
gradients.

We start with terms containing only the translational density and its gradients in Eq.\ \eqref{eq:Fexc_GpEa}.
In the vicinity of the smectic-A-isotropic transition one has for the smectic order parameter \cite{deGennes1973}
\begin{equation}
\psi(\vec{r}) = \psi_{0} e^{+\ii\varphi(\vec{r})}
\end{equation}
and for the density
\begin{equation}
\rho(\vec{r}) = \bar{\rho} + \psi_{0}[e^{+\ii\varphi(\vec{r})} + e^{-\ii\varphi(\vec{r})}]
\end{equation}
with the average homogeneous density $\bar{\rho}$
(compare also section 6.3 of reference \cite{ChaikinL1995} for a detailed discussion).
Since the total free energy must be a good scalar, the smectic order parameter can enter the free energy only quadratically.
For the lowest-order terms in the energy density $f(\vec{r})$, which we define as the integrand of the free-energy functional
\begin{equation}
\mathcal{F}[\rho,P,S] = \!\int_{\R^{2}} \:\!\!\!\!\!\!\dif\vec{r} \, f \;, 
\end{equation}
we have \cite{MukherjeePB2001}
\begin{equation}
\frac{1}{2} \alpha \lvert\psi\rvert^{2}
+\frac{1}{2} b_{1} \lvert\Nabla\psi\rvert^{2}
+\frac{1}{2} b_{2} \lvert\Laplace\psi\rvert^{2} \;.
\label{Aisohom}
\end{equation}
Comparing Eq.\ \eqref{Aisohom} and the first three terms in Eq.\ \eqref{eq:Fexc_GpEa},
we can make the identifications $A_{1}$, $A_{2}$, and $A_{3}$ with $-\alpha$, $-b_{1}$, and $-b_{2}$, respectively.
This situation is similar for non-polar nematics in three spatial dimensions \cite{WittkowskiLB2010},
where $A_{1}$, $A_{2}$, and $A_{3}$ are defined with different signs, and for non-polar nematics in two spatial dimensions \cite{Loewen2010}, 
where one must identify $4\pi^{2}\bar{\rho}A$, $-4\pi^{2}\bar{\rho}B$, and $4\pi^{2}\bar{\rho}\,C$ 
with $\alpha$, $b_{1}$, and $b_{2}$, respectively.

For the terms containing only the non-polar orientational order $S(\vec{r})$ in Eq.\ \eqref{eq:Fexc_GpEa}, 
we have two contributions to compare to other approaches. 
One is spatially homogeneous $\sim\!D_{1}$ and the other one is quadratic in the gradients of the orientational order $\sim\!D_{2}$.
The first contribution can be directly compared with the term $\frac{A}{2}Q_{ij}Q_{ij}$ in de Gennes' pioneering paper
\cite{deGennes1971}.
Using the structure $Q_{ij}=S(p_{i}p_{j}-\frac{1}{2}\delta_{ij})$ for the conventional nematic order parameter in two spatial dimensions, 
we find $D_{1}=-A$ using the original notation of reference \cite{deGennes1971}. 
For the gradient terms in the Ginzburg-Landau approximation one has at first sight two contributions to the energy density 
just using the three-dimensional expression \cite{deGennes1971}
\begin{equation}
L_{1}(\nabla_{i}Q_{jk})(\nabla_{i}Q_{jk})
+L_{2} (\nabla_{i}Q_{ik})(\nabla_{j}Q_{jk})
\end{equation}
for two spatial dimensions. A straightforward calculation shows that the two contributions are in two dimensions identical, however, with 
$L_{1}=2L_{2}$ and thus one independent coefficient just as for the PFC model where one has the contribution $\sim\!D_{2}$.

For the terms associated exclusively with orientational order we have, when specialized to two spatial dimensions,
in the continuum description in the energy density
\begin{equation}
\begin{split}
&K_{1}(\Nabla\cdot\hat{p})^{2} + K_{3}(\hat{p}\times [\Nabla\times\hat{p}])^{2} \\ 
+ \,& L_{\parallel}(p_{i}\nabla_{i}S)^{2}
+ L_{\perp}\delta^{\bot}_{ij}(\nabla_i S) (\nabla_j S) \\
+ \,& M (\nabla_i S) [\delta^{\bot}_{ik}p_{j} + \delta^{\bot}_{jk}p_{i}](\nabla_{j}p_{k}) \;,
\end{split}
\label{nemgrad}
\end{equation}
where $\delta^{\bot}_{ij}=\delta_{ij}-p_{i}p_{j}$ is the transverse Kronecker symbol projecting onto the direction 
perpendicular to the preferred direction $\hat{p}(\vec{r})$.
In Eq.\ \eqref{nemgrad}, the first line is connected to gradients of the director field $\hat{p}(\vec{r})$. 
It contains in two spatial dimensions only splay and bend and no twist and goes back to Frank's pioneering paper 
\cite{Frank1958,deGennesP1995}.
Lines 2 and 3 in Eq.\ \eqref{nemgrad} are associated with gradients of the nematic modulus, $S(\vec{r})$, and with a coupling term $\sim\!M$ 
between gradients of the director and gradients of the modulus \cite{BrandK1986,BrandP1987}.
We finally note that the gradient terms in Eq.\ \eqref{eq:Fexc_GpEa} are identical to the ones given in reference \cite{Loewen2010}, 
where we must identify $-D_{2}/2$ in the present paper with $2\pi^{2}\bar{\rho}E$ in reference \cite{Loewen2010}.
This must indeed be the case, since polar nematics contain the case of non-polar nematics as a special case in the PFC approach.

Next, we come to the terms containing only contributions of the macroscopic polarization $\vec{P}(\vec{r})$, or equivalently,
its magnitude (modulus) $P(\vec{r})$ and its direction $\hat{p}(\vec{r})$. 
The term $\sim\!C_{1}$ in Eq.\ \eqref{eq:Fexc_GpEa} is the standard quadratic term for a Landau expansion near, for example, 
the paraelectric-ferroelectric transition \cite{Kittel1995}. 
It also emerges when the phase transition isotropic to polar nematic is studied in Ginzburg-Landau approximation \cite{PleinerB1989}.
The terms that are quadratic in gradients of $\vec{P}(\vec{r})$, i.\ e., the contributions $\sim\!C_{2}$ and $\sim\!C_{3}$ in 
Eq.\ \eqref{eq:Fexc_GpEa}, can be compared to the result of a Ginzburg-Landau approach 
\begin{equation}
\tilde{D}_{1}(\nabla_{i}P_{i})(\nabla_{j}P_{j})
+\tilde{D}_{2}(\nabla_{i}P_{j})(\nabla_{i}P_{j})
\label{Pginz}
\end{equation}
and contain two independent contributions even in the isotropic phase \cite{PleinerB1989} in two spatial dimensions as is easily checked explicitly.

The gradient terms for the macroscopic polarization, or equivalently, for its magnitude $P(\vec{r})$ and its direction $\hat{p}(\vec{r})$, 
can also be compared to the macroscopic description of polar nematics \cite{BrandPZ2006,BrandCP2009}. 
For the corresponding terms we have
\begin{equation}
\begin{split}
& \frac{1}{2}K^{(2)}_{ij}(\nabla_{i}\delta P)(\nabla_{j}\delta P)
+ \frac{1}{2}K_{ijkl}(\nabla_{i}p_{j})(\nabla_{k}p_{l}) \\
&\!+ K^{(3)}_{ijk}(\nabla_{i}\delta P)(\nabla_{j}p_{k}) \;,
\end{split}
\label{Ginzmacro}
\end{equation}
where $\delta$ denotes deviations from the equilibrium value, in 
particular $\delta P=P-P_{0}$ 
and where the tensors are of the form
\begin{align}
\begin{split}
K_{ijkl} &= \frac{1}{2} K_{1}\big(\delta^{\bot}_{ij}\delta^{\bot}_{kl} + \delta^{\bot}_{il}\delta^{\bot}_{jk}\big) \\ 
&\quad\, + K_{3}\,p_{i}p_{k}\delta^{\bot}_{jl} \;, 
\end{split} \label{K} \\
\begin{split}
K^{(2)}_{ij} &= K_{4}\,p_{i}p_{j} + K_{5}\,\delta^{\bot}_{ij} \;, 
\end{split} \label{K2} \\
\begin{split}
K^{(3)}_{ijk} &= K_{6}\big(p_{i}\delta^{\bot}_{jk} + p_{j}\delta^{\bot}_{ik}\big) \;. 
\end{split} \label{K3}
\end{align}
Eq.\ \eqref{Ginzmacro} represents the analogue of the Frank orientational elastic energy ($\sim\!K_{ijkl}$) with splay and bend,
the energy associated with gradients of the modulus ($\sim\!K^{(2)}_{ij}$), and a cross-coupling term between gradients of the
preferred direction to gradients of the order-parameter modulus ($\sim\!K^{(3)}_{ijk}$) -- the analogue of the corresponding
term in non-polar nematics \cite{BrandK1986,KawasakiB1985}. 

The contributions $\sim\!C_{2}$ and $\sim\!C_{3}$ in Eq.\ \eqref{eq:Fexc_GpEa} are the PFC analogues of the contributions 
$\sim\!K^{(2)}_{ij}$ and $\sim\!K_{ijkl}$ in Eq.\ \eqref{Ginzmacro}.
Instead of four independent coefficients in the macroscopic description in two spatial dimensions, the PFC model gives rise to two. 
The contribution $\sim\!K_{6}$ has no direct analogue in the PFC model.

Next, we start to compare cross-coupling terms between gradients of the variables.
The discussion for the coupling terms between gradients of the density and gradients of the orientational order closely 
parallels that for the three-dimensional non-polar nematic case.
In Eq.\ \eqref{eq:Fexc_GpEa}, the terms of interest are proportional to $B_{3}$.
In reference \cite{WittkowskiLB2010}, these are the terms $\sim\!B_{2}$.
A comparison of these two expressions reveals that they are identical in structure and that one has just to take into account 
the change in dimensionality.
For spatial gradients in the director field coupling to spatial variations in the density $\rho(\vec{r})$ we find in the energy density 
\cite{PleinerB1980,BrandP1987}
\begin{equation}
\lambda^{\rho}(\nabla_{i}\rho)
[\delta^{\bot}_{ik}p_{j} + \delta^{\bot}_{jk}p_{i}](\nabla_{j}p_{k}) \;.
\label{ncoup}
\end{equation}
By comparison with Eq.\ \eqref{eq:Fexc_GpEa} we find $\lambda^{\rho}\bar{\rho}=B_{3}S$.
Finally, we have for the terms coupling gradients of the order-parameter modulus $S(\vec{r})$ to gradients of the density \cite{BrandP1987}
\begin{equation}
N^{\rho}_{ij}(\nabla_{i}S)(\nabla_{j}\rho) \;,
\label{Scoup}
\end{equation}
where the second rank tensor $N^{\rho}$ is of the standard uniaxial form
$N^{\rho}_{ij}=N^{\rho}_{1}p_{i}p_{j}+N^{\rho}_{2}\delta^{\bot}_{ij}$.
A comparison with Eq.\ \eqref{eq:Fexc_GpEa} yields $2N^{\rho}_{1}\bar{\rho}=B_{3}$ and 
$2N^{\rho}_{2}\bar{\rho}=-B_{3}$. 
The coupling terms listed in Eqs.\ \eqref{ncoup} and \eqref{Scoup} exist in both two and three spatial dimensions. 
Thus, in comparison to the hydrodynamic description of the bulk behavior, which is characterized by three independent coefficients, 
we find one independent coefficient in the PFC model. 
In the framework of a Ginzburg-Landau approach using the orientational order parameter $Q_{ij}(\vec{r})$ we find in the isotropic phase
\begin{equation}
P^{\xi}(\nabla_{i}Q_{jk})(\nabla_{l}\rho) 
(\delta_{ij}\delta_{kl}+\delta_{ik}\delta_{jl}) 
\label{Qrhoiso}
\end{equation}
and thus one independent coefficient -- as has also been the case for the non-polar PFC model in three dimensions \cite{WittkowskiLB2010}
as well as in two dimensions \cite{Loewen2010}.

The contributions $\sim\!B_{1}$ and $\sim\!B_{2}$ are containing gradients of the macroscopic polarization $\vec{P}(\vec{r})$ and couple to 
density and quadrupolar order.
They are unique to systems with polar order, or more generally, to systems with broken parity symmetry, since they contain one gradient and 
one factor $\vec{P}(\vec{r})$. Such coupling terms are not possible, for example, in non-polar nematics or smectic A phases.
The term $\sim\!B_{1}$ can easily be compared with the macroscopic description of polar nematics given in reference \cite{BrandPZ2006}. 
The relevant terms from Eq.\ (1) of reference \cite{BrandPZ2006} read 
\begin{equation}
\beta_{1}\delta\rho(p_{i}\nabla_{i}\delta P) 
+ \bar{\beta}_{1}\delta\rho(\nabla_{j}p_{j}) \;,
\label{beta1}
\end{equation}
where $\delta\rho = \rho - \bar{\rho}$.
We thus read off immediately that when comparing to the PFC model we have $2\beta_{1}\bar{\rho}=-B_{1}$ and 
$2\bar{\beta}_{1}\bar{\rho}=-B_{1}P$, that is 
one independent coefficient in the PFC model and two in the macroscopic description.
For the term $\sim\!B_{2}$ the situation is similar.
One has to replace in Eq.\ \eqref{beta1} $\delta\rho$ by $\delta S$, where $S(\vec{r})$ is the modulus of the quadrupolar
nematic order parameter with coefficients denoted by $\beta_{4}$ and $\bar{\beta}_{4}$. 
Then one makes the identifications $2\beta_{4}=-B_{2}$ and $2\bar{\beta}_{4}=B_{2}P$.
For the contribution $\sim\!B_{2}$ we can also make easily contact with the Ginzburg-Landau picture.
For the coupling of $P_{i}(\vec{r})$ and its gradients to quadrupolar orientational order we obtain to lowest order
in the Ginzburg-Landau energy density
\begin{equation}
g_{ijkl} P_{l}(\nabla_{k}Q_{ij}) 
\label{PQ}
\end{equation}
with $g_{ijkl}=g(\delta_{ik}\delta_{jl}+\delta_{il}\delta_{jk})$. 
This term has been given before for the isotropic-smectic-C$^{*}$ phase transition in liquid crystals \cite{MukherjeePB2005} 
for which the polarization $P_{i}(\vec{r})$ is a secondary-order parameter.
We note that the contribution $\sim\!B_{2}$ in Eq.\ \eqref{eq:Fexc_GpEa} can be brought into a form identical to that of 
Eq.\ \eqref{PQ}, when it is rewritten in terms of $Q_{ij}(\vec{r})$ and $\vec{P}(\vec{r})$.
This shows once more the close structural connection between PFC modeling and the Ginzburg-Landau approach. 

The spatially homogeneous contributions in Eq.\ \eqref{eq:Fexc_GpEb} can all be interpreted in the symmetry-based framework as well.
The term $\sim\!E_{4}$ arises near the smectic-C$^{*}$-isotropic phase transition \cite{MukherjeePB2005}: $Q_{ij}P_{i}P_{j}$.
The terms $\sim\!E_{2}$ and $\sim\!E_{3}$ can be interpreted as the density dependence of the terms $\sim\!\vec{P}^2$ and 
$\sim\!Q_{ij}Q_{ij}$ in the Landau description of the polar nematic-isotropic \cite{PleinerB1989} and the non-polar nematic-isotropic 
\cite{deGennes1971} phase transitions. Finally, the contribution $\sim\!E_{1}$ would arise in a macroscopic description as a term 
cubic in the density variations: $(\delta\rho)^3$. 
Typically, such terms are considered to be of higher order in a macroscopic approach. 
The physical interpretation of this term is a density dependence of the compressibility. 

Most of the terms in Eq.\ \eqref{eq:Fexc_GpEb} containing one gradient, namely all terms containing $F_{i}$, except for $F_{4}$,
can be interpreted in the framework of macroscopic dynamics as higher-order corrections to the terms $\sim\!\beta_{1}$, 
$\sim\!\beta_{4}$, $\sim\!\bar{\beta}_{1}$, and $\sim\!\bar{\beta}_{4}$ discussed above. 
They correspond in this picture to the dependence of the coefficients $\beta_{i}$ and $\bar{\beta}_{i}$ on the density changes
$\delta\rho(\vec{r})$ and variations in the modulus of the quadrupolar order parameter $\delta S(\vec{r})$. 
There is one exception to this picture and this is the term $\sim\!F_{4}$ in Eq.\ \eqref{eq:Fexc_GpEb}. 
It is also this term, which has an analogue in the field of the Ginzburg-Landau description of ferroelectric materials:
\begin{equation}
P_{i}P_{i}(\nabla_{j}P_{j}) \;.
\label{P2divP}
\end{equation}
This nonlinear gradient term has been introduced in reference \cite{AslanyanL1978} and it was demonstrated by 
Felix et al.\ \cite{FelixMH1986} that this term leads to qualitative changes in the phase diagram near the paraelectric-ferroelectric transition 
giving rise also to incommensurate structures.

In Eq.\ \eqref{eq:Fexc_GpEc}, spatially homogeneous terms that are of fourth order in the order parameters are presented. 
Most of them are familiar from Landau energies near phase transitions. The first contribution, the term $\sim\!G_{1}$, arises for all isotropic-smectic 
phase transitions \cite{MukherjeePB2001,MukherjeePB2002,MukherjeePB2005} as well as for the nematic-smectic-A and the nematic-smectic-C transitions 
\cite{deGennes1973,deGennesP1995}: $\sim\!\lvert\psi\rvert^{4}$.
The contribution $\sim\!G_{6}$ arises near the paraelectric-ferroelectric phase transition \cite{Kittel1995,FelixMH1986} and has also been used 
near the isotropic-polar-nematic transition \cite{PleinerB1989}: $\sim\!\vec{P}^{4}$.
The term $\sim\!G_{7}$ is familiar from the non-polar nematic to isotropic \cite{deGennes1971} and the smectic A to isotropic \cite{MukherjeePB2001} 
transitions: $\sim\!(Q_{ij}Q_{ij})^{2}$. 
The cross-coupling term $\sim\!G_{3}$ corresponds to an analogous term for isotropic-smectic transitions
\cite{MukherjeePB2001,MukherjeePB2002,MukherjeePB2005}: $\lvert\psi\rvert^{2}Q_{ij}Q_{ij}$. 
For the Ginzburg-Landau description of the smectic-C$^{*}$-isotropic transition, the term $\sim\!G_{2}$ arises \cite{MukherjeePB2005}: 
$\lvert\psi\rvert^{2}\vec{P}^{2}$.
The term $\sim\!G_{5}$ has also an analogue at the smectic-C$^{*}$-isotropic transition, where it has not been discussed before.
However, for the non-polar nematic to isotropic phase transition in an electric field one has shown in reference \cite{Brand1986}
that there are two contributions:
\begin{equation}
\tilde{\chi}_{1}E_{k}E_{n}Q_{kl}Q_{nl} 
+ \tilde{\chi}_{2}E_{n}E_{n}Q_{kl}Q_{kl} \;.
\end{equation}
The same contributions are relevant here when the external electric field is replaced by the polarization $\vec{P}(\vec{r})$.
Finally, the term $\sim\!G_{4}$ can be viewed as the density dependence of the term $Q_{ij}P_{i}P_{j}$ as it emerges near the
isotropic-smectic-C$^{*}$ phase transition \cite{MukherjeePB2005}.

\section{\label{sec:conclusions}Conclusions and possible extensions}
In conclusion, we systematically derived a phase-field-crystal model for polar liquid crystals in two spatial dimensions 
from microscopic density functional theory. Two basic approximations are involved: first, the density functional 
is approximated by a truncated functional Taylor expansion which we considered here up to fourth order. 
Then a generalized gradient expansion in the order parameters is performed which leads to a local 
free-energy functional. The density is parameterized by four order-parameter fields, namely the translational 
density $\psi_{1}(\vec{r})$ which corresponds to the scalar phase-field variable in the traditional phase-field-crystal model, 
the strength of polarization $P(\vec{r})$, an orientational direction given by a two-dimensional unit vector $\hat{p}$, 
and the nematic order parameter $S(\vec{r})$. In the three latter quantities, the gradient expansion is performed 
up to second order, while it is done to fourth order in $\psi_{1}(\vec{r})$ for stability reasons.
The traditional phase-field-crystal model \cite{ElderKHG2002,ElderG2004} and the recently proposed phase-field-crystal model 
for apolar liquid crystals \cite{Loewen2010} are recovered as special cases. The additional terms are all in accordance 
with macroscopic approaches based on symmetry considerations \cite{PleinerB1996,BrandP1987}. 
The prefactors are generalized moments of various direct correlation functions and therefore provide a bridge between microscopic 
and macroscopic approaches.

As a general feature, we find that typically the number of independent coefficients 
for the phase-field-crystal and the Ginzburg-Landau approaches is the same, while in many cases
the macroscopic hydrodynamics description valid inside the two-dimensional polar phase 
leads to a larger number of independent coefficients. This appears to be a general trend, 
which was also found to hold before for the comparison of phases with three-dimensional non-polar 
orientational order \cite{WittkowskiLB2010}.
In fact, it also applies to the two-dimensional phase-field-crystal model for systems with orientational order 
studied in reference \cite{Loewen2010}.

The proposed functional, as embodied in Eqs.\ \eqref{eq:Fexc_GpEa}-\eqref{eq:Fexc_GpEc}, can be used to study
phenomenologically phase transformations, for example, in polar nematic sheets, interfaces between coexisting phases
\cite{McDonaldAS2001,vanderBeekEtAl2006,BierHD2005}, and certain biological systems that exhibit polar order 
\cite{VerkhovskySB1999,CisnerosDGK2006}.
Since our model has more parameters, we expect even more complicated phase diagrams than recently numerically discovered
in the apolar phase-field-crystal model \cite{AchimWL2011}.

One could also do in principle microscopic calculations of the bulk phase diagram for a given interparticle potential
$\operatorname{V}(\vec{r}_{1}-\vec{r}_{2},\hat{u}_{1},\hat{u}_{2})$ which needs the full direct correlations 
of the isotropic phase as an input. The simplest idea is to neglect all direct correlation functions for $n\geqslant 3$ and to rely 
on a second-order virial expression \cite{vanRoijBMF1995}, where 
$c^{(2)}(\vec{r}_{1}-\vec{r}_{2},\hat{u}_{1},\hat{u}_{2})=e^{-\beta\operatorname{V}(\vec{r}_{1}-\vec{r}_{2},\hat{u}_{1},\hat{u}_{2})}-1$, 
or the random-phase approximation for mean-field fluids \cite{RexWL2007}, where
$c^{(2)}(\vec{r}_{1}-\vec{r}_{2},\hat{u}_{1},\hat{u}_{2})=-\beta\operatorname{V}(\vec{r}_{1}-\vec{r}_{2},\hat{u}_{1},\hat{u}_{2})$.

In a next step, the analysis can be done for Brownian dynamics based on dynamical density functional theory 
\cite{MariniMT1999,ArcherE2004,EspanolL2009}, which was generalized to 
orientational dynamics \cite{RexL2008} and can be used as a starting point to derive the order-parameter dynamics 
\cite{Loewen2010}. This can then be applied to describe the translational and orientational relaxation dynamics, 
for example, for an orientational glass \cite{RennerLB1995} or system exposed to a periodic driving field \cite{HaertelBL2010}.
Finally, it would be interesting to generalize the analysis to self-propelled particles which are driven along 
their orientation \cite{TonerTR2005,Ramaswamy2010,PeruaniDB2006}. These particles are polar by definition and therefore 
the generalization to dynamics of the present theory is mandatory to derive microscopic theories \cite{WensinkL2008,ElgetiG2009} 
for their collective swarming behavior.
A dynamical theory could for example be used to investigate the dynamical properties of bacterial growth patterns of 
\emph{proteus mirabilis} \cite{WatanabeEtAl2002}.

\begin{acknowledgments}
We thank Michael Schmiedeberg and Yuka Tabe for helpful discussions.
This work has been supported by the Deutsche Forschungsgemeinschaft within SPP 1296.
H.R.B. thanks the Deutsche Forschungsgemeinschaft for partial support of his work through the 
Forschergruppe FOR 608 "Nichtlineare Dynamik komplexer Kontinua".
\end{acknowledgments}

\appendix
\section{\label{Anhang:SC}Symmetry considerations}
In the derivation of the approximation for the excess free-energy functional, a large number of expansion coefficients
$\tilde{c}^{(n)}_{\underline{l},\underline{m}}(\underline{R})$ of the direct correlation functions and moments 
$\mathrm{M}^{\underline{m}}_{\underline{l}}(\underline{\alpha})$ of these expansion coefficients appear. 
To reduce their total number, we used basic symmetry considerations that base on \emph{four invariances} of the direct correlation functions and 
showed that many of the expansion coefficients and moments are equal. This is why only a few moments of all possible moments for different 
index combinations are present in the equations \eqref{eq:Ca}-\eqref{eq:Cg} for the coefficients in our model.  
These invariances are the \emph{translational and rotational invariance} of the direct correlation functions, which are considered by an appropriate 
parametrization $c^{(n+1)}(\underline{R},\underline{\phi_{\mathrm{R}}},\underline{\phi})$ and a Fourier expansion \eqref{eq:cexp} of the latter, 
as well as the invariance of the direct correlation functions concerning the \emph{renumbering of particles},
\begin{equation}
\begin{split}
&c^{(n)}(\dotsc,\vec{r}_{i},\dotsc,\vec{r}_{j},\dotsc,\dotsc,\hat{u}_{i},\dotsc,\hat{u}_{j},\dotsc) \\
=\,&c^{(n)}(\dotsc,\vec{r}_{j},\dotsc,\vec{r}_{i},\dotsc,\dotsc,\hat{u}_{j},\dotsc,\hat{u}_{i},\dotsc)\;,
\end{split}
\end{equation}
which implies that moments that arise from each other by simultaneous permutations of the elements in 
$\underline{l}$, $\underline{m}$, and $\underline{\alpha}$ are equal,
\begin{equation}
\begin{split}
&\mathrm{M}^{\dotsc,m_{i},\dotsc,m_{j},\dotsc}_{\dotsc,l_{i},\dotsc,l_{j},\dotsc}
(\dotsc,\alpha_{i},\dotsc,\alpha_{j},\dotsc) \\
=\,&\mathrm{M}^{\dotsc,m_{j},\dotsc,m_{i},\dotsc}_{\dotsc,l_{j},\dotsc,l_{i},\dotsc}
(\dotsc,\alpha_{j},\dotsc,\alpha_{i},\dotsc) \;,
\end{split}
\end{equation}
and the invariance of the expansion coefficients \eqref{eq:ec} against \emph{complex conjugation}:
\begin{equation}
\overline{\tilde{c}^{(n)}_{\underline{l},\underline{m}}(\underline{R})}=\tilde{c}^{(n)}_{\underline{l},\underline{m}}(\underline{R}) \;.
\end{equation}
The last assumption is necessary to obtain physical terms with real coefficients in the approximation for the excess free-energy functional. 
It involves the invariance of $\tilde{c}^{(n)}_{\underline{l},\underline{m}}(\underline{R})$ against simultaneous reversal of the signs of the 
elements in $\underline{l}$ and $\underline{m}$,
\begin{equation}
\begin{split}
& \tilde{c}^{(n)}_{-l_{1},\dotsc,-l_{n},\,-m_{1},\dotsc,-m_{n}}(R_{1},\dotsc,R_{n})\\
=\,&\tilde{c}^{(n)}_{l_{1},\dotsc,l_{n},\,m_{1},\dotsc,m_{n}}(R_{1},\dotsc,R_{n}) \;,
\end{split}
\end{equation}
and is equivalent to the invariance of the direct correlation functions against reflection of 
the system at the first axis of coordinates.

When the system is apolar, the liquid-crystalline particles have head-tail symmetry. In this case, the modulus $P(\vec{r})$ of the polarization
is zero and its orientation $\hat{p}(\vec{r})$ is not defined, while the direction $\hat{n}(\vec{r})$ associated with quadrupolar order still exists.
Then, further symmetry considerations lead to the following equalities between 
expansion coefficients of the direct pair-correlation function: 
\begin{equation}
\begin{split}
\tilde{c}^{(2)}_{-1,1}(R)&=\tilde{c}^{(2)}_{1,0}(R)\;,\\
\tilde{c}^{(2)}_{-1,2}(R)&=\tilde{c}^{(2)}_{1,1}(R)\;,\\
\tilde{c}^{(2)}_{-2,2}(R)&=\tilde{c}^{(2)}_{2,0}(R)\;.
\end{split}
\end{equation}
The consequence of these equations is, that the coefficients $B_{1}$ and $B_{2}$ vanish and $B_{3}$ becomes more simple.

\bibliography{References}
\end{document}